\documentclass[preprint]{aastex63}
\submitjournal{ApJL}
\shorttitle{Solar prominence formation via supergranulation}
\shortauthors{Liu et al.}
\begin{document}

\title{Formation of Quiescent Prominence Magnetic Field by Supergranulations}

\correspondingauthor{Chun Xia}
\email{chun.xia@ynu.edu.cn}

\author{Qingjun Liu}
\affiliation{School of Physics and Astronomy, Yunnan University \\
 Kunming 650500, China}

\author{Chun Xia}
\affiliation{School of Physics and Astronomy, Yunnan University \\ 
 Kunming 650500, China}
\affiliation{National Astronomical Observatories, Chinese Academy of Sciences \\
 Beijing 100101, China}

\begin{abstract}

To understand the formation of quiescent solar prominences, the origin of their magnetic 
field structures, i.e., magnetic flux ropes (MFRs), must be revealed. We use three-dimensional
magnetofriction simulations in a spherical subdomain to investigate the role of typical supergranular motions 
in the long-term formation of a prominence magnetic field. Time-dependant horizontal 
supergranular motions with and without the effect of Coriolis force are simulated on the 
solar surface via Voronoi tessellation. The vortical motions by the Coriolis effect at 
boundaries of supergranules inject magnetic helicity into the corona. The helicity is 
transferred and accumulated along the polarity inversion line (PIL) as strongly sheared 
magnetic field via helicity condensation. The diverging motions of supergranules converge 
opposite magnetic polarities at the PIL and drive the magnetic reconnection between 
footpoints of the sheared magnetic arcades to form an MFR. The magnetic network, 
negative-helicity MFR in the northern hemisphere, and fragmented-to-continuous formation 
process of magnetic dip regions are in agreement with observations. Although diverging 
supergranulations, differential rotation, and meridional flows are included, the simulation 
without the Coriolis effect can not produce an MFR or sheared arcades to host a prominence.
Therefore Coriolis force is a key factor for helicity injection and the formation
of magnetic structures of quiescent solar prominences.

\end{abstract}


\section{Introduction} \label{sec:intro}

Quiescent prominences and filaments are long thin structures of dense and cool plasma 
that remain stable in the solar corona for days or weeks along the polarity inversion 
lines (PILs) of quiescent regions \citep{Parenti2014}. Their magnetic fields have special 
structures, so-called filament channels, which support the dense mass against gravity
\citep{Mackay2010}. The aligned chromospheric fibrils at two sides of a filament 
\citep{Foukal1971} and dominant magnetic field components along the PILs of prominences 
\citep{Leroy1983} indicate that filament channels have strongly sheared magnetic fields. 
The overall magnetic topology of quiescent filament channels is most likely a helical 
magnetic flux rope (MFR) \citep{Xia2014a,Xia2014b,Xia2016} with plenty of observational evidence,
such as the elliptical shape of coronal cavities around quiescent prominences 
\citep{Gibson2010}, concentric rings of Doppler velocity in coronal cavities \citep{Bak2013},
persistent swhirling motions of coronal plasma around the center of coronal cavities above 
prominences \citep{Wang2010}, and the inverse polarity of prominence magnetic field whose 
component perpendicular to the PIL is opposite to the one of a potential field 
\citep{Bommier1994}. Dipped sheared arcades with concave-up magnetic fields 
\citep{DeVore2000} may be the magnetic structure of short filament channels around active 
regions, but not long ones in quiescent regions \citep{Patsourakos2020}.

Two mechanisms were proposed to explain the formation of MFRs in the corona.
The first mechanism suggests that the emergence of the upper part of an MFR from the 
conversion zone into the corona leads to sheared magnetic arcades (SMAs), arcade-like magnetic loops
without concave-up magnetic fields, which are then transformed into
an MFR through magnetic reconnection \citep{Fan2001}. However, this flux emergence 
mechanism does not apply to the filaments in quiescent regions where no large-scale
magnetic flux emergence was found \citep{Mackay2008}. Statistical study of filament 
observations reveals that over 90\% of filaments lie above PILs external to 
conjugate bipolar fields by flux emergence \citep{Mackay2008}. The second mechanism 
relies on the shearing, converging, and cancellation of opposite-polarity magnetic 
flux on the photosphere, which transforms SMAs into a helical 
MFR by magnetic reconnection in the lower solar atmosphere \citep{vanBallegooijen1989}. 
This flux cancellation mechanism has been supported by many numerical simulations 
from magnetofriction models \citep{Mackay2006} and zero-beta magnetohydrodynamic (MHD) 
models \citep{Amari1999} to isothermal MHD models \citep{Xia2014a}. However, these models 
were confronted with three difficulties. First, these models use smooth magnetic 
flux distributions, which contradicts the observational fact that the photospheric magnetic flux 
under quiescent filaments is discrete and concentrated as small elements at supergranular 
boundaries \citep{Zhou2021}. Second, the models relied on large-scale systematic 
converging flows or equivalent magnetic flux diffusion towards the PIL on the 
photosphere to drive the flux cancellation. But observations have not found such large-scale 
flows, instead, supergranular-scale converging flows between diverging supergranules were 
found both inside and outside filament channels \citep{Rondi2007,Schmieder2014}. 
Third, the models required large-scale shearing motions, such as differential rotation, 
to get shear arcades. However, the differential rotation injects helicity of the opposite sign
at west-east oriented PILs in contrast to the observed hemispheric preference 
\citep{vanBallegooijen1998}, which manifests that filaments with negative (positive) 
helicity dominate in the northern (southern) hemisphere \citep{Ouyang2017}.

\citet{Antiochos2013} proposed the helicity condensation theory to explain the formation 
of filament channels. In the theory, photospheric vortical motions 
between the convective cells inject magnetic helicity into magnetic flux tubes
and coronal magnetic reconnections between neighboring flux tubes transfer 
the injected twist towards the outer periphery of the flux tubes, leading to an inverse 
cascade of magnetic helicity from small scales to the largest scale at the periphery of 
the whole magnetic flux system along the PIL, where the helicity condenses and SMAs appear. 
The theory was demonstrated by ideal MHD simulations based on topological coronal models between 
two parallel plates \citep{Zhao2015,Knizhnik2015} and a Cartesian coronal model with a circular PIL \citep{Knizhnik2017}.
These numerical models ignored the primary diverging flows of supergranular cells and simplified
 the observed cyclonic vortices between the anticyclonic flows of supergranules 
\citep{Duvall2000,Langfellner2015} caused by the Coriolis force \citep{Hathaway1982, Egorov2004}, 
as many annular vortical flows to inject magnetic helicity into corona and 
produced the SMAs along PILs. Using a large-scale averaged representation 
of the small-scale vortical motions in magnetofriction simulations, \citet{Mackay2014} 
found that the helicity condensation can overcome the incorrect sign of helicity injection
from differential rotation on a west-east oriented PIL and help to form MFRs in filament 
channels.
In this Letter, we investigate the role of supergranular flows in the formation 
of the magnetic field of quiescent prominences via magnetofriction simulations. 

\section{Numerical Methods} \label{sec:methods}

The simulation domain is a spherical subdomain extending over radii 
$r\in[1 R_\odot, 1.5 R_\odot]$, colatitudes $\theta\in[39.6^\circ,90^\circ]$, and 
longitudes $\phi\in[0^\circ,60^\circ]$ with a 4-level adaptive mesh refinement (AMR) to get $256^3$ 
effective resolution. From a smooth bipolar photospheric magnetogram, we use the spherical potential 
field extrapolation module in the PDFI\_SS software \citep{Fisher2020} to extrapolate the initial
potential magnetic field. We solve the magnetofriction equations \citep{Guo2016} without explicit 
resistivity using the code MPI-AMRVAC \citep{Porth2014,Xia2018}. Magnetic reconnection
here is caused by numerical resistivity.
The magnetofrictional velocity $v=\mathbf{J}\times\mathbf{B}/(\nu_0 B^2)$, where 
$\nu_0=10^{-15}$ s cm$^{-2}$ is the viscous coefficient, has a smooth decay to zero 
towards the photosphere \citep{Cheung2012} and an upper limit of $30$ km s$^{-1}$ 
\citep{Pomoell2019}. We use the constrained transport (CT) scheme \citep{Gardiner2005}, with 
the HLL flux for the electric field, on the staggered AMR mesh \citep{Olivares2019} to keep 
magnetic field divergence-free. Boundary conditions are periodic on the longitudinal 
boundaries, open on the outer radial boundary, and closed on the latitudinal boundaries. 
On the photospheric boundary, we use zero normal velocity with the supergranular 
horizontal velocity field described as follows.

A Voronoi tessellation of solar surface resembles (super)granular segmentation in 
both topology and statistics \citep{Schrijver1997}. We use a weight function 
$\omega=3|\sin(\pi t/\tau+\xi)|+0.7$ to generate Voronoi tessellation, in which $\tau$ is 
the lifetime of a supergranular cell with a Gaussian random distribution centered on 1.6 days 
\citep{Hirzberger2008} and $\xi$ is the random initial phase. We use 1 as the lower limit
of the weights for living cells, smaller weights are set to zero to remove the corresponding 
cells from the tessellation map. So the simulated supergranular cells are changing with
time with their area proportional to their weights. We adopt the dimensionless diverging velocity within
supergranular cells as $v_r(r)=2r^2/r_0 \exp(-4r^2/r_0^2)$ \citep{Gudiksen2005}, where $r$ is the spherical 
distance to the cell center and $r_0$ is the radius of a circle with area equal to the supergranular cell. 
The vortical velocity due to Coriolis force acting on the diverging horizontal flows is simplified
as $v_t(r)=2r/r_0 v_r(r)$ neglecting the latitude-dependence of Coriolis force. We use a normalization 
factor to make sure the maximal supergranular speed is initially 500 m s$^{-1}$. The
velocities of differential rotation and meridional flow \citep{Mackay2014} are also included. 
We multiply the total driving speed by 5 to speed up the long-term evolution. Dimensionless values 
in our results have the time unit of 8.3 hours and the magnetic field unit of 2 Gauss. We have run a 
model SR with supergranules rotating under Coriolis force and a model SS without the Coriolis effect.

\section{Results} \label{sec:results}

\begin{figure}[ht!]
\includegraphics[width=\textwidth]{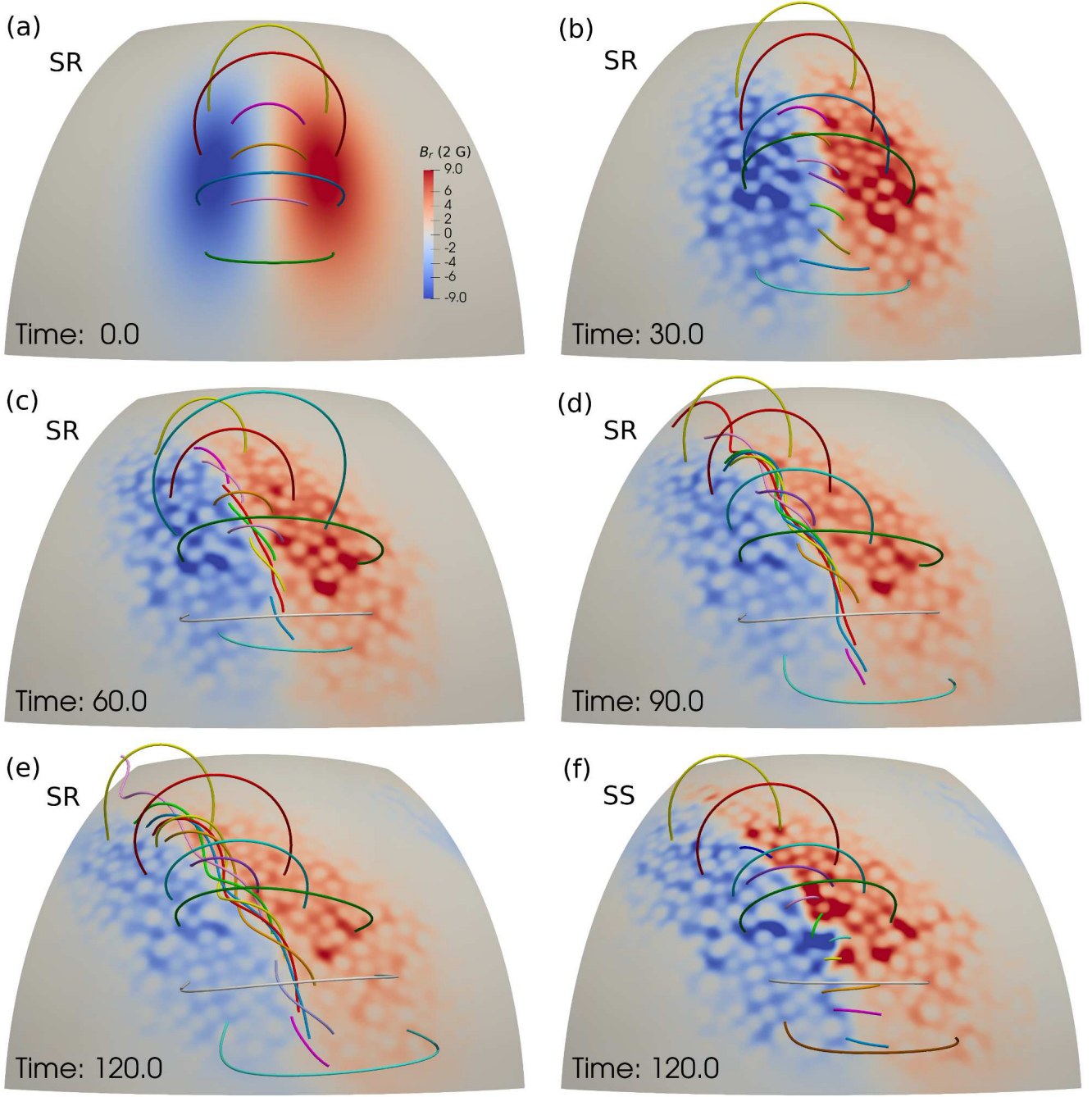}
\caption{Magnetic field lines of model SR with Coriolis effect at time 0 (a), 30 (b), 60 (c), 
90 (d), and 120 (e) in contrast to model SS without Coriolis effect at time 120 (f). The photosphere is 
colored by the radial magnetic field in red-to-blue colors saturated at $\pm$18 G.
\label{fig:evolve}}
\end{figure}

Initially (Figure~\ref{fig:evolve} (a)), the magnetic field is a bipolar arcade rooted in the smooth 
positive (red) and negative (blue) polarity regions separated by a straight north-south directed PIL. 
Then, the supergranular diverging flows transport and concentrate magnetic flux to the boundaries of 
supergranules forming a magnetic network, and the differential rotation gradually shifts the northern 
part of magnetic flux to the east and the southern part to the west making the PIL tilted anticlockwise. 
At time 30 (Figure~\ref{fig:evolve} (b)), short magnetic loops near the PIL have skewed away from the 
initial potential field state. The skew angle, defined as the angle between the horizontal component 
of the magnetic field above the PIL and the horizontal direction perpendicular to the PIL \citep{Mackay2001}, 
is about 45 degrees.  Long loops, which root in the central region of each polarity far from the PIL, 
remain in an approximately potential state with neglilible skew angles during the whole evolution. 
At time 60 (Figure~\ref{fig:evolve} (c)), low-lying magnetic loops near the PIL are 
almost aligned with the PIL with large skew angles and some of them are changed into helical field lines 
(red line) along the PIL. At time 90 (Figure~\ref{fig:evolve} (d)), more helical field lines form and 
enwind the earlier-born helical lines assembling a low-lying thin MFR along the PIL.  Later at time 120 
(Figure~\ref{fig:evolve} (e)), which is after 41.5 days, the MFR matures with a larger height and 
cross-section size suitable for hosting a quiescent prominence. Different from earlier MFR models in which 
helical field lines root in two compact regions, the footpoints of this MFR distribute extensively along the 
 PIL. In contrast to model SR, model SS does not produce any MFR or SMAs until time 120 
(Figure~\ref{fig:evolve} (f)), and all magnetic loops stay nearly potential for another 120 time.

\begin{figure}[ht!]
\includegraphics[width=6.in]{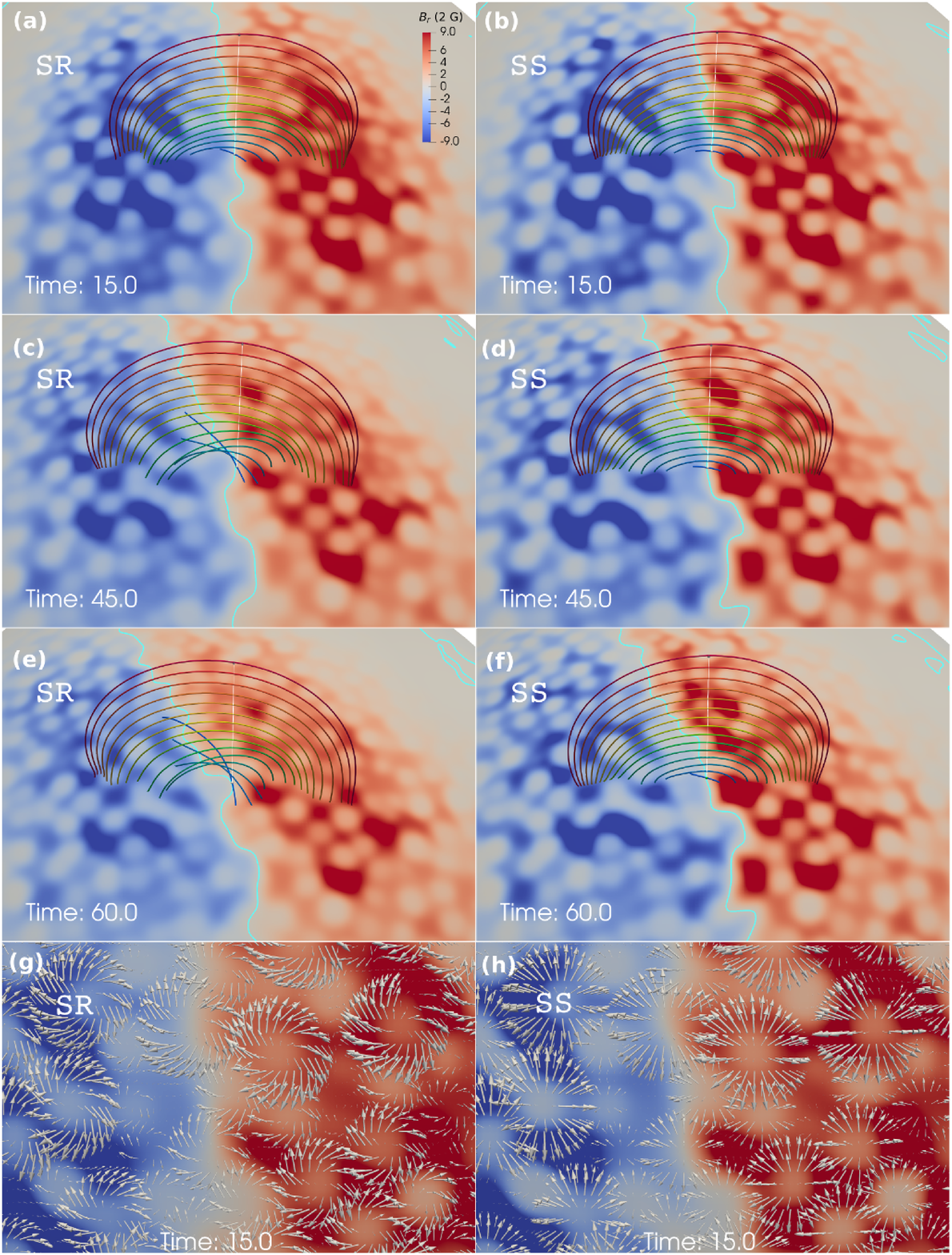}
\caption{Comparative views of model SR and SS at time 15, 45, and 60. In panel (a)-(f), magnetic field lines,
in rainbow colors, are integrated from uniformly sampled seed points along a radial white line starting 
from the cyan PIL at a fixed latitude. The photosphere is colored by radial magnetic field 
saturated at $\pm$20 G. Zoom-in views at time 15 are shown in panel (g) and (h), where grey arrows present 
the photospheric horizontal velocity field. 
\label{fig:shear}}
\end{figure}

To understand the origin of the sheared magnetic loops, we illustrate the build-up process of 
the sheared loops of model SR in the left column of Figure~\ref{fig:shear}. For comparison, a similar 
representation of model SS at the same time is plotted in the right column. In panels (a)-(f), the 
magnetic loops crossing over the cyan PIL at different heights are plotted in rainbow colors. 
At time 15, the lowest blue loop in model SR has a short length and a small skew angle to 
the PIL, while the loops in model SS have nearly zero skew angles. For model SR at later times 45 
and 60, several loops close to the PIL are sheared and the lowest loop has the fastest growth rate 
in the skew angle and length. Further away from the PIL, loops with higher altitudes have smaller 
skew angles. In contrast, loops in model SS at all heights and distances from the PIL, remain 
roughly perpendicular to the PIL. The sheared loops with large skew angles in model SR are not 
created by persistent photospheric shearing flows, because such shearing flows along the PIL do 
not exist. Instead, only vortical diverging and converging flows by supergranules exist as shown 
by grey arrows in panels (g) and (h). A logical explanation of the strongly sheared loops around 
the PIL in model SR is referable to the helicity condensation theory \citep{Antiochos2013}.
Vortical converging flows towards the junctions of several supergranules inject negative 
magnetic helicity into the magnetic flux tubes with footpoints at strong magnetic flux on the 
photosphere. Component magnetic reconnection happens on the interface of adjacent twisted flux tubes, 
which transports the twisted flux to the common periphery of the flux tubes and leaves behind the 
untwisted flux in the central regions. This process inversely cascades from small scales to large 
scales until reaching the biggest flux tube representing 
the whole bipolar region with part of the periphery folded around the PIL, where twisted fluxes with 
negative helicity condense as SMAs.

\begin{figure}[ht!]
\includegraphics[width=0.6\textwidth]{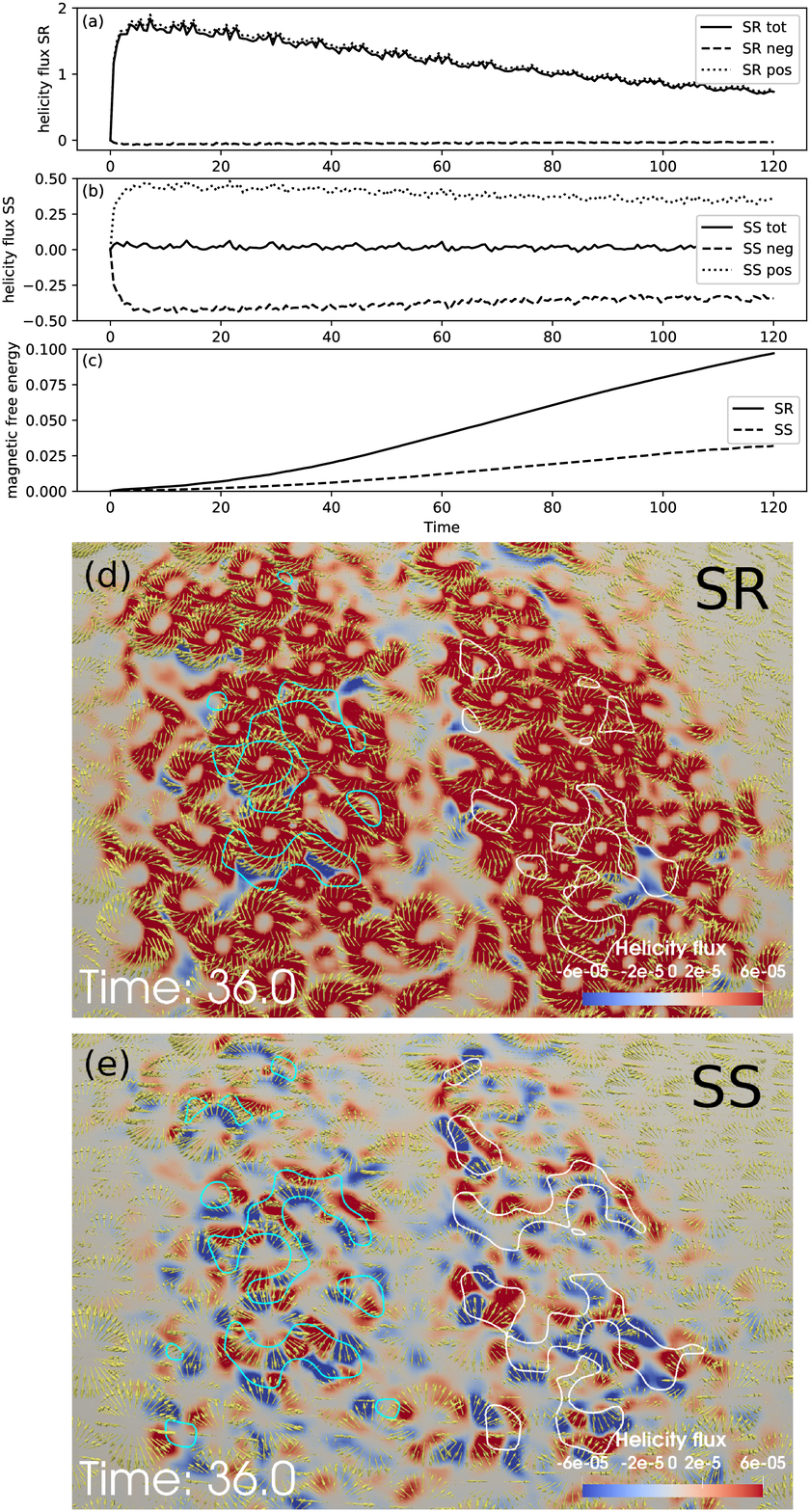}
\caption{(a) Time evolution of the magnetic helicity fluxes as the total (solid line), the negative 
(dashed line), and the positive (dotted line) ones through the photosphere of model SR. 
(b) Similar curves as (a) for model SS. (c) Time evolution of the magnetic free energy of model SR (solid line) 
and SS (dashed line). (d) Magnetic helicity flux density on the photosphere overlaid with horizontal velocity field 
as yellow arrows and contours of $B_r$ at $\pm16$ G for model SR at time 36. (e) Similar plot as (d) for model SS.
\label{fig:tevolution}}
\end{figure}

Then we present how much the magnetic helicity is injected into the corona and how the 
magnetic free energy is stored in our models. The flux of (relative) magnetic helicity
through the photosphere is defined by the equation with zero resistivity \citep{Berger1984}:
\begin{equation}\label{eqn1}
\frac{d H_R}{dt}=-2\int \mathbf{A}^P\times\mathbf{E}\cdot d\mathbf{S}=-2\int (A^P_\theta E_\phi-A^P_\phi E_\theta)dS
\end{equation}
where $\mathbf{A}^P$ is the vector potential of the potential magnetic field which matches the instantaneous photospheric $B_r$,
$\mathbf{E}$ is the electric field and quantified as the electric field from the CT scheme, and $d\mathbf{S}$ is 
the surface element with normal direction $\mathbf{\hat{r}}$. Figure~\ref{fig:tevolution} (a) shows the time evolution of the 
magnetic helicity fluxes through the photosphere in model SR as the total (solid line), the negative 
(dashed line), and the positive flux (dotted line). The positive helicity flux is about 20 to 30 times larger 
than the negative helicity flux. Magnetic helicity fluxes through other boundaries are negligible.
The photospheric distribution of the integrand in equation (\ref{eqn1}) is shown in 
Figure~\ref{fig:tevolution} (d) with horizontal velocity vectors and contours of $B_r$ overplotted. The positive 
values of the integrand are distributed in the outer regions of supergranules in donut shapes, induced by the clockwise 
rotating diverging flows of supergranules. The negative values are distributed sporadically along the borders 
of several supergranules and caused by irregular counterclockwise converging flows towards strong magnetic flux.
For model SS as shown in panel (b), the positive and negative helicity flux almost cancel out with a small positive 
total flux. Panel (e) for model SS shows that positive values alternates with negative values in the outer region of
 supergranules. Figure~\ref{fig:tevolution} (c) presents the time evolution of the magnetic free energy, which is obtained by 
subtracting the potential field energy from the total magnetic energy. Both models have an increasing 
magnetic free energy, while the magnetic free energy in model SR increases about three times faster than it 
in model SS.

\begin{figure}[ht!]
\includegraphics[width=\textwidth]{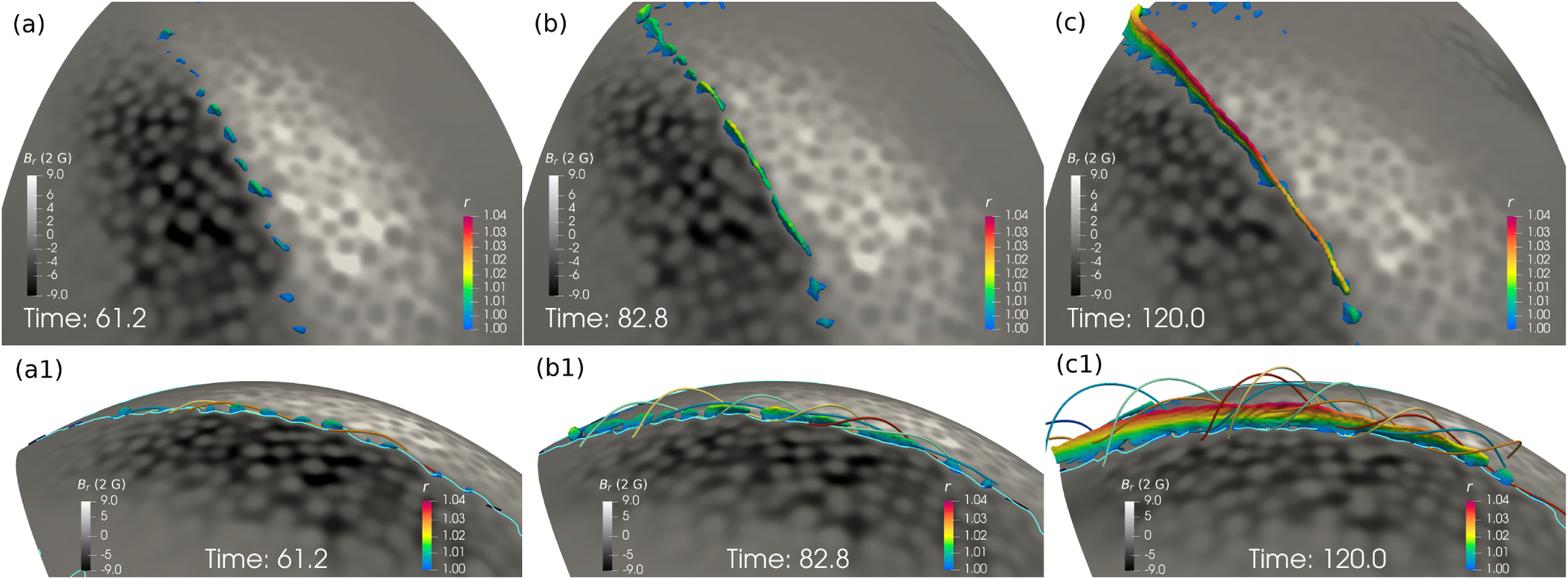}
\caption{Formation of magnetic dip regions shown as isosurfaces colored by solar radius $r$ from top views 
(a)-(c) and from side views (a1)-(c1) at time 61.2, 82.8, and 120. Surrounding magnetic field lines in 
different colors and the main PIL in cyan are plotted in (a1)-(c1) additionally.
\label{fig:dipevolution}}
\end{figure}

Long-lived prominence plasma should be supported by magnetic tension force against downward gravity in 
locally concave-up magnetic fields, i.e., magnetic dips. We locate the magnetic dip regions (MDRs) where the 
radial component of the curvature of magnetic field lines is positive and the radial component of the magnetic 
field has less than 10\% proportion. Figure~\ref{fig:dipevolution} presents snapshots of the formation of 
magnetic dips from the top views in upper panels (a)-(b) and from the side views in lower pannels (a1)-(b1).
At time 61.2, small isolated MDRs distribute along the PIL with supergranular scale intervals. Later at 
time 82.8, the small MDRs grow horizontally along the PIL and connect to form three long MDRs. Helical 
field lines tangential to the photosphere outline the periphery of an MFR above the PIL and the MDRs 
naturally appear in the lower half of the MFR. At time 120, further growth and connection of MDRs result 
in a long slab-like MDR with a growing height reaching 28,000 km and a length of about 480,000 km. The MFR 
surrounding the MDR has helical field lines winding 1 to 2 turns around a common axis. If the MDRs are filled
with prominence plasma, the formation process of MDRs is consistent with filament observations showing 
that quiescent filaments form from aligned fragments to a continuous body \citep{Pevtsov2005}.

\begin{figure}[ht!]
\includegraphics[width=\textwidth]{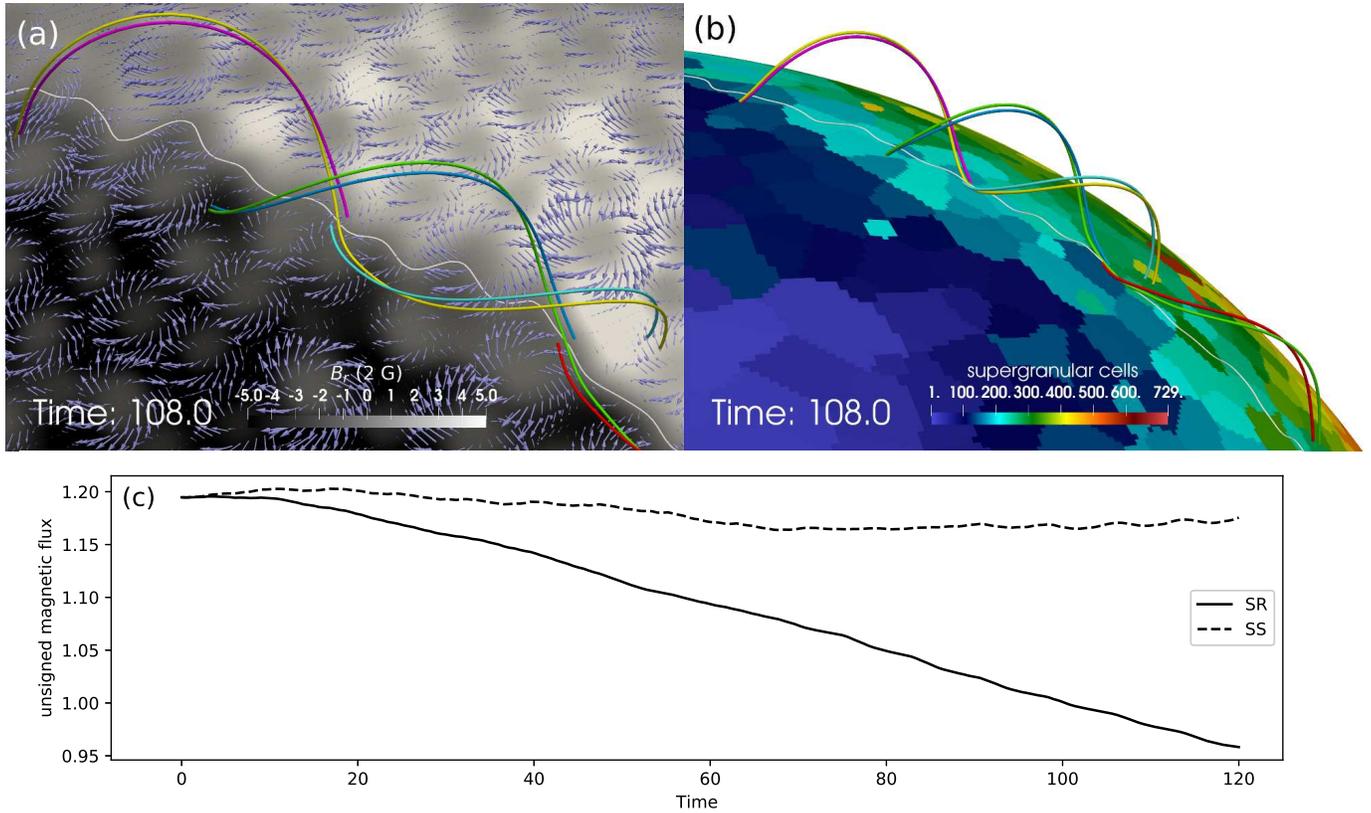}
\caption{Linking sheared loops to helical field lines via photospheric magnetic reconnection
along the white PIL. (a) Top view of 6 selected magnetic field lines above the photospheric 
magnetogram with arrows presenting horizontal velocity. (b) Side view of the same magnetic field lines 
in (a) above a photospheric color map presenting supergranular cells. (c) Evolution of total unsigned 
magnetic flux on photosphere for SR (solid line) and SS (dashed line) model.
\label{fig:cancellation}}
\end{figure}

To understand the transition from SMAs to MFRs, we find the sites of 
magnetic reconnection and flux cancellation on the photosphere. In Figure~\ref{fig:cancellation} (a),
the footpoints of the pink and the sky blue field line, in the middle of the panel, converge 
to the white PIL by adjacent supergranular flows indicated by the purple arrows, and the two field lines
are about to be reconnected at the PIL to form a helical field line like the yellow one. Similarly, the 
blue and the red field line will be reconnected to form a helical field line like the green one in the lower
right region. Figure~\ref{fig:cancellation} (b) presents a zoom-out side view of these field lines 
above the photospheric supergranular cells. The sites of magnetic reconnection are at the borders of 
supergranular cells on two sides of the PIL. Some parts of the PIL crossing through supergranular cells 
are bald patches where the MFR touches the photosphere. 
Evidence of magnetic flux cancellation is shown in Figure~\ref{fig:cancellation} (c)
in which the time evolution of the total unsigned magnetic flux on the photosphere is plotted for both models.
After a short period of flux dispersion without many cancellations, the unsigned magnetic flux of model SR 
decreases linearly losing about 20 \% until time 120, which indicates continuous flux cancellation. The 
unsigned magnetic flux of model SS decreases less than 3 \% until time 70 and then slightly increases 
with small oscillation and slow flux cancellation. The slight increase is caused by numerical error
of the magnetic boundary condition.

\section{Discussion and Conclusions} \label{sec:conclusion}

The only difference in model setup between model SR and SS is the vortical motions by Coriolis force, 
which leads to strikingly different results, namely, an MFR in model SR and potential arcades in 
model SS. The counterclockwise vortical flows in strong magnetic flux regions around supergranular 
boundaries inject negative magnetic helicity, which is accumulated along the PIL via helicity condensation
 to form a dextral filament channel in the northern hemisphere. Therefore Coriolis force is a key 
factor for helicity injection and formation of quiescent prominences, while effective magnetic flux 
diffusion by supergranular flows \citep{Leighton1964} and differential rotation are not the main 
reasons for the formation of filament channels. We will include the latitude dependence of the Coriolis
effect \citep{Duvall2000} in the future expecting faster MFR formation at higher latitudes. 

The helicity condensation theory descibes how the injected magnetic helicity at small scale
 is transferred and condensed to the PIL, which explain the origin of strong axial magnetic flux 
along quiescent filament channels. But this theory can only lead to SMAs, which contradict the 
observational evidence \citep{Bak2013,Wang2010} of quiescent prominences. Previous 
numerical simulations on helicity condensation \citep{Zhao2015,Knizhnik2015,Knizhnik2017} did not produce MFRs 
because of the absent of the primary diverging motions of supergranules, which not only generate the 
magnetic network but also converge opposite magnetic polarities at the PIL and drive the magnetic 
reconnection between footpoints of SMAs, the magnetic flux cancellation in forming filament 
channels \citep{Wang2007}, and the formation of MFRs.

During the formation of the coherent MFR in model SR, a chain of small MFRs first forms along the PIL with 
fragmented small MDRs to host prominence plasma. Later on, these small MFRs and MDRs grow and merge with 
neighbors to form the mature MFR, which is consistent with the ``head-to-tail" conceptual model of prominence 
formation proposed by \citet{Martens2001}. The isolated MDRs consist of piled-up magnetic dips from the photosphere
may correspond to the observed pillars or legs of quiescent prominences \citep{Li2013,Zhou2021}. We run 
models starting from west-to-east PILs to simulate high-latitude quiescent prominences and formed MFRs to 
be reported in a follow-up paper.

\acknowledgments

This research was supported by the Basic Research Program of Yunnan Province (2019FB140, 202001AW070011), 
the National Natural Science Foundation of China (11803031, 12073022). We acknowledge discussions with 
Mark C. M. Cheung, George H. Fisher, and Shangbin Yang. The numerical simulations were conducted on the 
Yunnan University Astronomy Supercomputer.

\software{MPI-AMRVAC, PDFI\_SS}


\begin{thebibliography}{}
\expandafter\ifx\csname natexlab\endcsname\relax\def\natexlab#1{#1}\fi
\providecommand{\url}[1]{\href{#1}{#1}}
\providecommand{\dodoi}[1]{doi:~\href{http://doi.org/#1}{\nolinkurl{#1}}}
\providecommand{\doeprint}[1]{\href{http://ascl.net/#1}{\nolinkurl{http://ascl.net/#1}}}
\providecommand{\doarXiv}[1]{\href{https://arxiv.org/abs/#1}{\nolinkurl{https://arxiv.org/abs/#1}}}

\bibitem[{Amari {et~al.}(1999)Amari, Luciani, Mikic, \& Linker}]{Amari1999}
Amari, T., Luciani, J.~F., Mikic, Z., \& Linker, J. 1999, The Astrophysical
  Journal Letters, 518, L57

\bibitem[{Antiochos(2013)}]{Antiochos2013}
Antiochos, S.~K. 2013, The Astrophysical Journal, 772, 72,
  \dodoi{10.1088/0004-637X/772/1/72}

\bibitem[{Bak-SteAlicka {et~al.}(2013)Bak-SteAlicka, Gibson, Fan, Bethge,
  Forland, \& Rachmeler}]{Bak2013}
Bak-SteAlicka, U., Gibson, S.~E., Fan, Y., {et~al.} 2013, The Astrophysical
  Journal, 770, L28, \dodoi{10.1088/2041-8205/770/2/L28}

\bibitem[{Berger \& Field(1984)}]{Berger1984}
Berger, M.~A., \& Field, G.~B. 1984, Journal of Fluid Mechanics, 147, 133,
  \dodoi{10.1017/S0022112084002019}

\bibitem[{Bommier {et~al.}(1994)Bommier, Degl'Innocenti, Leroy, \&
  Sahal-Brechot}]{Bommier1994}
Bommier, V., Degl'Innocenti, E.~L., Leroy, J.~L., \& Sahal-Brechot, S. 1994,
  Solar Physics, 154, 231

\bibitem[{Cheung \& DeRosa(2012)}]{Cheung2012}
Cheung, M. C.~M., \& DeRosa, M.~L. 2012, The Astrophysical Journal, 757, 147,
  \dodoi{10.1088/0004-637X/757/2/147}

\bibitem[{DeVore \& Antiochos(2000)}]{DeVore2000}
DeVore, C.~R., \& Antiochos, S.~K. 2000, The Astrophysical Journal, 539, 954

\bibitem[{Duvall \& Gizon(2000)}]{Duvall2000}
Duvall, T.~L., \& Gizon, L. 2000, Solar Physics, 192, 177,
  \dodoi{10.1007/978-94-011-4377-6_10}

\bibitem[{Egorov {et~al.}(2004)Egorov, Rudiger, \& Ziegler}]{Egorov2004}
Egorov, P., Rudiger, G., \& Ziegler, U. 2004, Astronomy and Astrophysics, 425,
  725, \dodoi{10.1051/0004-6361:20040531}

\bibitem[{Fan(2001)}]{Fan2001}
Fan, Y. 2001, The Astrophysical Journal Letters, 554, L111

\bibitem[{Fisher {et~al.}(2020)Fisher, Kazachenko, Welsch, Sun, Lumme, Bercik,
  DeRosa, \& Cheung}]{Fisher2020}
Fisher, G.~H., Kazachenko, M.~D., Welsch, B.~T., {et~al.} 2020, The
  Astrophysical Journal Supplement Series, 248, 2,
  \dodoi{10.3847/1538-4365/ab8303}

\bibitem[{Foukal(1971)}]{Foukal1971}
Foukal, P. 1971, Solar Physics, 19, 59

\bibitem[{Gardiner \& Stone(2005)}]{Gardiner2005}
Gardiner, T.~A., \& Stone, J.~M. 2005, Journal of Computational Physics, 205,
  509, \dodoi{10.1016/j.jcp.2004.11.016}

\bibitem[{Gibson {et~al.}(2010)Gibson, Kucera, Rastawicki, Dove, de~Toma, Hao,
  Hill, Hudson, MarquÃ©, McIntosh, Rachmeler, Reeves, Schmieder, Schmit,
  Seaton, Sterling, Tripathi, Williams, \& Zhang}]{Gibson2010}
Gibson, S.~E., Kucera, T.~A., Rastawicki, D., {et~al.} 2010, The Astrophysical
  Journal, 724, 1133, \dodoi{10.1088/0004-637X/724/2/1133}

\bibitem[{Gudiksen \& Nordlund(2005)}]{Gudiksen2005}
Gudiksen, B.~V., \& Nordlund, A. 2005, The Astrophysical Journal, 618, 1020

\bibitem[{Guo {et~al.}(2016)Guo, Xia, Keppens, \& Valori}]{Guo2016}
Guo, Y., Xia, C., Keppens, R., \& Valori, G. 2016, The Astrophysical Journal,
  828, 82, \dodoi{10.3847/0004-637X/828/2/82}

\bibitem[{Hathaway(1982)}]{Hathaway1982}
Hathaway, D.~H. 1982, Solar Physics, 77, 341.
\newblock \url{http://articles.adsabs.harvard.edu/pdf/1982SoPh...77..341H}

\bibitem[{Hirzberger {et~al.}(2008)Hirzberger, Gizon, Solanki, \&
  Duvall}]{Hirzberger2008}
Hirzberger, J., Gizon, L., Solanki, S.~K., \& Duvall, T.~L. 2008, Solar
  Physics, 251, 417, \dodoi{10.1007/s11207-008-9206-8}

\bibitem[{Knizhnik {et~al.}(2015)Knizhnik, Antiochos, \& DeVore}]{Knizhnik2015}
Knizhnik, K.~J., Antiochos, S.~K., \& DeVore, C.~R. 2015, The Astrophysical
  Journal, 809, 137, \dodoi{10.1088/0004-637X/809/2/137}

\bibitem[{Knizhnik {et~al.}(2017)Knizhnik, Antiochos, DeVore, \&
  Wyper}]{Knizhnik2017}
Knizhnik, K.~J., Antiochos, S.~K., DeVore, C.~R., \& Wyper, P.~F. 2017, The
  Astrophysical Journal, 851, L17, \dodoi{10.3847/2041-8213/aa9e0a}

\bibitem[{Langfellner {et~al.}(2015)Langfellner, Gizon, \&
  Birch}]{Langfellner2015}
Langfellner, J., Gizon, L., \& Birch, A.~C. 2015, Astronomy \& Astrophysics,
  581, A67, \dodoi{10.1051/0004-6361/201526024}

\bibitem[{Leighton(1964)}]{Leighton1964}
Leighton, R.~B. 1964, The Astrophysical Journal, 140, 1547,
  \dodoi{10.1086/148058}

\bibitem[{Leroy {et~al.}(1983)Leroy, Bommier, \& Sahal-Brechot}]{Leroy1983}
Leroy, J.~L., Bommier, V., \& Sahal-Brechot, S. 1983, Solar Physics, 83, 135

\bibitem[{Li \& Zhang(2013)}]{Li2013}
Li, L., \& Zhang, J. 2013, Solar Physics, 282, 147,
  \dodoi{10.1007/s11207-012-0122-6}

\bibitem[{Mackay {et~al.}(2014)Mackay, DeVore, \& Antiochos}]{Mackay2014}
Mackay, D.~H., DeVore, C.~R., \& Antiochos, S.~K. 2014, The Astrophysical
  Journal, 784, 164, \dodoi{10.1088/0004-637X/784/2/164}

\bibitem[{Mackay {et~al.}(2008)Mackay, Gaizauskas, \& Yeates}]{Mackay2008}
Mackay, D.~H., Gaizauskas, V., \& Yeates, A.~R. 2008, Solar Physics, 248, 51,
  \dodoi{10.1007/s11207-008-9127-6}

\bibitem[{{Mackay} {et~al.}(2010){Mackay}, {Karpen}, {Ballester}, {Schmieder},
  \& {Aulanier}}]{Mackay2010}
{Mackay}, D.~H., {Karpen}, J.~T., {Ballester}, J.~L., {Schmieder}, B., \&
  {Aulanier}, G. 2010, \ssr, 151, 333, \dodoi{10.1007/s11214-010-9628-0}

\bibitem[{Mackay \& van Ballegooijen(2001)}]{Mackay2001}
Mackay, D.~H., \& van Ballegooijen, A.~A. 2001, The Astrophysical Journal, 560,
  445, \dodoi{10.1086/322385}

\bibitem[{Mackay \& Van~Ballegooijen(2006)}]{Mackay2006}
Mackay, D.~H., \& Van~Ballegooijen, A.~A. 2006, The Astrophysical Journal, 641,
  577

\bibitem[{Martens \& Zwaan(2001)}]{Martens2001}
Martens, P.~C., \& Zwaan, C. 2001, The Astrophysical Journal, 558, 872

\bibitem[{Olivares {et~al.}(2019)Olivares, Porth, Davelaar, Most, Fromm,
  Mizuno, Younsi, \& Rezzolla}]{Olivares2019}
Olivares, H., Porth, O., Davelaar, J., {et~al.} 2019, Astronomy \&
  Astrophysics, 629, A61, \dodoi{10.1051/0004-6361/201935559}

\bibitem[{Ouyang {et~al.}(2017)Ouyang, Zhou, Chen, \& Fang}]{Ouyang2017}
Ouyang, Y., Zhou, Y.~H., Chen, P.~F., \& Fang, C. 2017, The Astrophysical
  Journal, 835, 94, \dodoi{10.3847/1538-4357/835/1/94}

\bibitem[{Parenti(2014)}]{Parenti2014}
Parenti, S. 2014, Living Reviews in Solar Physics, 11,
  \dodoi{10.12942/lrsp-2014-1}

\bibitem[{Patsourakos {et~al.}(2020)Patsourakos, Vourlidas, Torok, Kliem,
  Antiochos, Archontis, Aulanier, Cheng, Chintzoglou, Georgoulis, Green, Leake,
  Moore, Nindos, Syntelis, Yardley, Yurchyshyn, \& Zhang}]{Patsourakos2020}
Patsourakos, S., Vourlidas, A., Torok, T., {et~al.} 2020, Space Science
  Reviews, 216, 131, \dodoi{10.1007/s11214-020-00757-9}

\bibitem[{Pevtsov \& Neidig(2005)}]{Pevtsov2005}
Pevtsov, A.~A., \& Neidig, D. 2005, in {ASP} {Conference} {Series}, Vol. 346,
  219--226.
\newblock \url{http://adsabs.harvard.edu/abs/2005ASPC..346..219P}

\bibitem[{Pomoell {et~al.}(2019)Pomoell, Lumme, \& Kilpua}]{Pomoell2019}
Pomoell, J., Lumme, E., \& Kilpua, E. 2019, Solar Physics, 294, 41,
  \dodoi{10.1007/s11207-019-1430-x}

\bibitem[{Porth {et~al.}(2014)Porth, Xia, Hendrix, Moschou, \&
  Keppens}]{Porth2014}
Porth, O., Xia, C., Hendrix, T., Moschou, S.~P., \& Keppens, R. 2014, The
  Astrophysical Journal Supplement Series, 214, 4,
  \dodoi{10.1088/0067-0049/214/1/4}

\bibitem[{Rondi {et~al.}(2007)Rondi, Roudier, Molodij, Bommier, Keil,
  SÃ¼tterlin, Malherbe, Meunier, Schmieder, \& Maloney}]{Rondi2007}
Rondi, S., Roudier, T., Molodij, G., {et~al.} 2007, Astronomy \& Astrophysics,
  467, 1289, \dodoi{10.1051/0004-6361:20066649}

\bibitem[{Schmieder {et~al.}(2014)Schmieder, Roudier, Mein, Mein, Malherbe, \&
  Chandra}]{Schmieder2014}
Schmieder, B., Roudier, T., Mein, N., {et~al.} 2014, Astronomy \& Astrophysics,
  564, A104, \dodoi{10.1051/0004-6361/201322861}

\bibitem[{Schrijver {et~al.}(1997)Schrijver, Hagenaar, \&
  Title}]{Schrijver1997}
Schrijver, C.~J., Hagenaar, H.~J., \& Title, A.~M. 1997, The Astrophysical
  Journal, 475, 328

\bibitem[{van Ballegooijen {et~al.}(1998)van Ballegooijen, Cartledge, \&
  Priest}]{vanBallegooijen1998}
van Ballegooijen, A.~A., Cartledge, N.~P., \& Priest, E.~R. 1998, The
  Astrophysical Journal, 501, 866, \dodoi{10.1086/305823}

\bibitem[{Van~Ballegooijen \& Martens(1989)}]{vanBallegooijen1989}
Van~Ballegooijen, A.~A., \& Martens, P. C.~H. 1989, The Astrophysical Journal,
  343, 971

\bibitem[{Wang \& Muglach(2007)}]{Wang2007}
Wang, Y.-M., \& Muglach, K. 2007, The Astrophysical Journal, 666, 1284

\bibitem[{Wang \& Stenborg(2010)}]{Wang2010}
Wang, Y.-M., \& Stenborg, G. 2010, The Astrophysical Journal, 719, L181,
  \dodoi{10.1088/2041-8205/719/2/L181}

\bibitem[{Xia \& Keppens(2016)}]{Xia2016}
Xia, C., \& Keppens, R. 2016, The Astrophysical Journal, 823, 22,
  \dodoi{10.3847/0004-637X/823/1/22}

\bibitem[{Xia {et~al.}(2014{\natexlab{a}})Xia, Keppens, Antolin, \&
  Porth}]{Xia2014b}
Xia, C., Keppens, R., Antolin, P., \& Porth, O. 2014{\natexlab{a}}, The
  Astrophysical Journal, 792, L38, \dodoi{10.1088/2041-8205/792/2/L38}

\bibitem[{Xia {et~al.}(2014{\natexlab{b}})Xia, Keppens, \& Guo}]{Xia2014a}
Xia, C., Keppens, R., \& Guo, Y. 2014{\natexlab{b}}, The Astrophysical Journal,
  780, 130, \dodoi{10.1088/0004-637X/780/2/130}

\bibitem[{Xia {et~al.}(2018)Xia, Teunissen, Mellah, Chane, \&
  Keppens}]{Xia2018}
Xia, C., Teunissen, J., Mellah, I.~E., Chane, E., \& Keppens, R. 2018, The
  Astrophysical Journal Supplement Series, 234, 30,
  \dodoi{10.3847/1538-4365/aaa6c8}

\bibitem[{Zhao {et~al.}(2015)Zhao, DeVore, Antiochos, \& Zurbuchen}]{Zhao2015}
Zhao, L., DeVore, C.~R., Antiochos, S.~K., \& Zurbuchen, T.~H. 2015, The
  Astrophysical Journal, 805, 61, \dodoi{10.1088/0004-637X/805/1/61}

\bibitem[{Zhou {et~al.}(2021)Zhou, Xia, \& Shen}]{Zhou2021}
Zhou, C., Xia, C., \& Shen, Y. 2021, Astronomy \& Astrophysics, 647, A112,
  \dodoi{10.1051/0004-6361/202039558}

\end{thebibliography}
\end{document}